# Structure and Electrical Properties of DNA Nanotubes Embedded in Lipid Bilayer Membranes

Himanshu Joshi and Prabal K. Maiti*

Center for Condensed Matter Theory, Department of Physics, Indian Institute of Science,

Bangalore 560012, India

*To whom correspondence should be addressed. Tel: +091-80-22932865
Fax: +91-80-23602602; Email: maiti@iisc.ac.in

**Abstract**

Engineering the synthetic nanopores through lipid bilayer membrane to access the interior of a cell is a long persisting challenge in biotechnology. Here, we demonstrate the stability and dynamics of a tile-based 6-helix DNA nanotube (DNT) embedded in POPC lipid bilayer using the analysis of 0.2 µs long equilibrium MD simulation trajectories. We observe that the head groups of the lipid molecules close to the lumen cooperatively tilt towards the hydrophilic sugar-phosphate backbone of DNA and form a toroidal structure around the patch of DNT protruding in the membrane. Further, we explore the effect of ionic concentrations to the in-solution structure and stability of the lipid-DNT complex. Transmembrane ionic current measurements for the constant electric field MD simulation provide the I-V characteristics of the water filled DNT lumen in lipid membrane. With increasing salt concentrations, the measured values of transmembrane ionic conductance of the porous DNT lumen vary from 4.3 nS to 20.6 nS. Simulations of the DNTs with ssDNA and dsDNA overhangs at the mouth of the pore show gating effect with remarkable difference in the transmembrane ionic conductivities for open and close state nanopores.

**INTRODUCTION**

Lipid bilayer membranes define the boundaries of a biological cell and regulate various kinds of cellular transport across it. The assembly of lipid molecules in aqueous medium is governed by the amphiphilic interactions where the polar head groups face outside and shield the hydrophobic tails from water to form a bilayer. Various peptides self-assemble to form pores which allow access of the interior of cell through these impermeable bilayers. These pores can be further engineered for various biophysical applications and synthetic biology.(1-4) Over past three decades, structural DNA nanotechnology has emerged as a convenient approach to creating nanostructures of arbitrary shape with sub-nanometer precision.(5-8) Several experimental groups have recently shown that self-assembled DNA nanostructures can mimic the naturally occurring nanopores in lipid bilayer membrane.(9-12) The diameter of these DNA Nanopores (DNPs) varies from 1.6 nm to 4 nm.(13-15) Due to highly functionalized and automated chemical properties along with the conformational polymorphism, DNPs are easier to customize as compared to the conventional protein nanopores. In order to compensate the energy cost of pore formation in membrane, the charged sugar phosphate backbone of membrane spanning region of DNA is modified by covalently conjugating the hydrophobic lipid anchors like ethyl phosphorothioate,(10) cholesterol,(11,14,16) streptavidin,(15) porphyrin(17) etc. DNPs in membrane have been characterized through advanced imaging techniques like confocal fluorescent microscopy, atomic force microscopy (AFM), aragose gel electrophoresis and single ion channel current recordings. These findings have injected fresh aspirations in field of synthetic nanopores. Recent experiments have shown that these nanopores are capable of the selective transport of molecular cargo across in membrane.(15,16) The versatile self-assembled DNPs can be further useful for diverse biotechnological applications like targeted drug delivery, biosensing, genome sequencing and tools for biophysical studies etc.(18) Seifert et al. showed a voltage dependent gating of 6-helix DNP in lipid bilayer membranes associated with two states of conductance.(19) With the advancements in atomistic and coarse-grained representations of DNA, MD simulations have become a handy tool to predict the properties of self-assembled DNA nanostructures.(20-24) Yoo and Aksimentiev have studied several types of lipid spanning DNPs using atomistic and coarse grained MD simulation which revealed several inherent aspects of these DNPs like ionic conductance, mechanical gating, electro-osmatic pressure effects etc.(25,26) The assembly of DNA and lipid is very rare in nature due to their contrasting features and interactions. It is imperative to study the interaction between DNA and lipid membranes for their rational design.(27) Aiming to understand the microscopic structure and the molecular interaction governing the self-assembly of DNPs in ambient conditions, herein we present an all atom molecular dynamics (MD) simulation study of DNT embedded in lipid bilayer membrane. We have created an atomistic model of experimentally characterized tile based 6-helix DNT (6HB) with an inner diameter of 2 nm, embedded in POPC lipid bilayer membrane. This DNT has been experimentally synthesized by Wang *et al.* (28) and we have investigated the nanoscale structure and elasticity of the same using the MD simulations.(29) As it has also been observed that the nature of counterions and salt concentration strongly affects the structure and stability of nucleic acids (30) as well as the behavior of lipid bilayer membranes,(31) we also investigate the stability of these DNTs in lipid membrane as a function of monovalent ($Na^+$ and $K^+$) salt concentrations. Moreover, we measure the transmembrane ionic current and the Ohmic conductance of the channel from the MD simulations at constant electric field. Based on the analysis of multiple MD simulation trajectories, we propose a novel mechanism of the reorientation of lipid head groups to form a toroidal structure around DNT lumen in lipid bilayer membrane. This mechanism provides stability to the structure of DNPs in lipid bilayer membrane by reducing the energy barrier of insertion. We find that the structure of the DNT is most stable in 1M NaCl salt solution. We expect that our simulation study with atomistic resolution will provide important information about the physical properties of DNPs in lipid membrane and will guide future design of DNPs in this burgeoning field which is still in its infancy from various aspects.

## METHODS

**Computational Design of DNT in POPC Lipid Bilayer Membrane**

There are two widely known approaches to generate DNTs, namely tiles based self-assembly(32) and DNA origami techniques. It has been known that due to the simplicity of the construction methods, higher yields and lower molecular weight, DNTs assembled from crossover tiles are easy to customize further to assemble as membranes nanopore.(13) These DNTs are essentially the extension of DX(33) and TX(34) crossover molecules into a tubular geometry. We built atomistic model of 57 base pair (bp) long DNT using a custom-built NAB programme. The NAB code operates with "base first" strategy, then creating the sugar-phosphate backbone and connecting the phosphodiester bond according to the design. Six dsDNA, kept at the vertices of a hexagon of length 2 nm, were interconnected with adjacent dsDNA at two crossovers points on each side. These crossovers are Holliday like junctions, where the strands of DNA switch its double helical domain.(35) The distance between two such crossovers is 7 bp or integer multiples of 7 bp. The design of the crossovers and sequence of nucleotides of DNT are taken from the experimental work by Wang et al.(28) Previously, we have extensively investigated the *in situ* structure, stability and elasticity of this DNT using molecular simulation.(29,36) 1-palmitoyl 2-oleoyl-sn-glycero3-phosphocholine (POPC) is mono-cis unsaturated zwitterionic phospholipid with two asymmetric hydrocarbon chains, P and O, attached to phosphatialdylcholines (PCs). We generated the initial configuration of the POPC lipid bilayer membrane using CHARMM-GUI membrane builder.(37) Next, we created a pore in the membrane by removing all the lipid molecules within a radius of 3 nm from the center of the membrane as shown in top panel of Figure S1a in supporting information (SI). The DNT was inserted in the membrane pore by aligning the helical axis of DNT and bilayer normal in such a way that the center of mass of the membrane and DNT matches as shown in Figure 1a and bottom panel of Figure S1a in SI. The method of generating the nanopore and insertion of DNT is similar to the recent study on cytolysin-A (cly-A) pore in POPC lipid bilayer membrane by Mandal et al.(38) We solvated the DNT embedded lipid bilayer membrane in a rectangular water box using TIP3P water model.(39) Here, we ensured 20 Å water shell from the outermost atom of DNA along the DNT axis (z axis). Monovalent counterions (Na+/K+) were added to neutralize the negative charge of the DNT. Further, we added additional Na+/K+ and Cl- ions to create the systems with four different molar concentrations i.e. 0.5M NaCl, 1M NaCl and 0.5M KCl, 1M KCl. The leap module creates a Coulombic grid of 1 Å around the solute and places the ions at lowest electrostatic potential. During the process of ion placement, some overlapping water molecules were replaced by ions. Figure S1b in SI shows the side view of the initial built structure of DNT embedded in lipid membrane immersed into explicit salt solution. Table S1 summarizes the atomistic details of all the five simulated systems.

**Simulation Methodology**

All atom MD simulation have been performed using AMBER14 simulation package.(40) We used ff99(41,42) force field parameters with parmbsc0(43) and the recent $\epsilon\zeta$ dihedral refinements(44) to describe the bonded and non-bonded interactions for DNA. Lipid14 force field parameters were used to describe the interaction involving the POPC lipid molecules.(45) We used Joung-Cheatham ion parameters optimized for TIP3P water model along with the Lorentz-Berthelot combination rule to describe their crucial non-bonded interaction in the system.(46) These force field parameters are extensively validated, are widely used in biomolecular simulations.(47) The built structures were subjected to a series of steepest descent and conjugated gradient energy minimization steps to remove any bad contacts in the system while slowly removing the harmonic restraints from the solute. The systems were gradually heated upto 300 K in 105 ps MD simulation with 1 fs time step. While heating, the lipid and DNA atoms were harmonically restraint with 20 kcal/mol.Å$^{-2}$ force constant. Subsequently, we equilibrated the system for 5 ns at 1 atm pressure and 300 K temperature to acquire the correct density. Finally, 205 ns production MD simulation were carried out in NPT ensemble using 2 fs integration time step and periodic boundary condition. The translation motion of the center of mass of the system was removed after every 1000 time steps. We used Langevin thermostat with 2 ps$^{-1}$ collision frequency and anisotropic pressure coupling with Berendsen barostat to maintain the constant temperature and pressure. We choose a new random number at the interval of 1 ns to avoid the synchronization artifacts in Langevin dynamics. Particle Mesh Ewald (PME) method integrated with AMBER simulation package was implemented to calculate the short range part of the Coulombic interaction as well as the long range electrostatic interactions between the various components of the system.(48) We used 10 Å cutoff to calculate the non-bonded interaction. All the bonds involving hydrogen were restrained using SHAKE algorithm.(49). To measure the ionic current as a function of transmembrane voltage difference, we performed MD simulation at constant electric field using NAMD code with AMBER force field parameters.(50) We have extensively used CPPTRAJ(51) and VMD(52) programming suites for various analysis and graphics presented in this study. We have previously characterized the *in silico* stability of various DNA based nanostructures using the similar protocol.(53-55)

## RESULTS AND DISCUSSIONS

### Tilt Angle of Lipids Chains: Lipid DNT Interaction

Hydrophilic nature of DNA backbone makes the insertion of the DNT into the hydrophobic core of lipid bilayer membrane energetically unfavorable. Experimentally, researchers have tried to overcome this problem by modifying the DNA backbone with hydrophobic lipid based anchors to compensate the energy loss. Closely following the snapshots of the simulation, we see that for the lipid molecules close to the pore, the lipid head groups cooperatively tilt towards the DNT and shield the tail molecules by forming a toroidal shape around the water filled DNT channel within first few ns of the simulation as shown in schematically on the top panel of figure 1. This conformation is maintained throughout the 205 ns long MD simulation. Figure 1 shows the instantaneous snapshots of the lipid head group atoms (Nitrogen and Phosphorus) at the beginning of the simulation, after 15 ns and at the end of 205 ns equilibrium MD simulation. Similar type of toroidal lipid structure was also observed in recent study by Gopfrich et al with funnel-shaped DNA origami nanopore.(14) We quantify the tilting of lipid molecules towards the nanopore by measuring their tilt angles. Here, the tilt angle of the lipid molecule is described by defining two tilt angles corresponding to palmitoyl (P) and oleoyl (O) alkylic chain of the POPC lipid molecule. The tilt angle is defined as

$$\arccos\left(\sqrt{cos^2\theta}\right)$$

where $\theta$ is the angle between bilayer normal and the average segmental vector of the respective chain as depicted in figure 2a. Here, the segment vector is the line joining n-1 to n+1 carbon atoms in the alkyl chain. Similar scheme is routinely used to calculate the tilt angle of free standing lipid bilayer membrane in MD simulations. Further, we classified the lipid molecules into two groups, near and far, depending on the distance of the corresponding head group atoms from the center of the pore. We used a 3 nm distance cutoff to define the near and far lipid as shown in figure 2b where the near lipid have been highlighted in blue color. We calculate the tilt angle for all the lipid molecules and average them to get a mean tilt angle. Figure 2c and 2d shows the time evolution of average tilt angles for P and O chains of the POPC respectively. These plots describe that the near lipids distinctly tilts towards the central pore by large measure giving rise to higher tilt angle whereas the far lipids largely remains in the initial conformation similar to the free standing lipid bilayer membrane. Similar behavior was found in all the simulations at different molar concentrations. Figure 2e and 2f represent the normalized probability distribution of the tilt angles of far and near lipid molecules averaged over the last 10 ns of the 200 ns long MD simulation for P and O chains respectively. For both the chains, the most probable tilt angle for near lipids is significantly larger as compared to the far lipids. The values of average tilt angle from the last 10 ns equilibrium MD simulation have been summarized in the Table 1. We have also encountered similar reorientation of lipid molecules in our previous electroporation study on POPC and DMPC lipid bilayers. (56) This phenomenon has also been observed by Gopfrich et al. in their combined experimental and computational studies through the simulation snapshots of the local density of lipid head groups on ion channels made from porphyrin tagged single duplex DNA.(25) The lipid reorientation mechanism is very much likely to play an important role in stability of the amphipathic DNT nanopore in lipid bilayer membrane. A similar mechanism was also suggested by Khalid et al. for DNA translocation through the lipid bilayer membrane.(57) To see any possible diffusion of DNT across the membrane, we performed additional 0.5 µs long simulation for 0 M system. As discussed in appendix S2 of SI, the analysis suggests that the DNT structure is preserved in membrane over longer time scale as well.

**Stability and Conformational Evolution**

The structure and stability of biomolecules are strongly dependent on the nature of counterions and molarity of salt solution.(58) We have compared the structure and conformational stability of DNT embedded in lipid membrane varying the molarity of the solution both for NaCl and KCl buffer solution. Figure 3 shows the instantaneous snapshots of system after 205 ns equilibrium MD simulations at various salt concentrations. These snapshots suggest that tubular structure of DNT is better maintained at higher molarity. We calculated the Root-Mean-Square-Deviation (RMSD) of DNT with respect to the initial energy minimized structure. Figure 4a compares the time evolution of the RMSD for DNT at different salt concentrations both for NaCl and KCl solution. The higher salt concentration gives better stability to the system. We find that the RMSD for the DNT structure is least for the 1M NaCl concentration followed by 1M KCl, 0.5NaCl, 0.5KCl and 0M respectively. We have also calculated the DNT radius profile at various salt concentrations and the radius profile is shown in Figure 4b. To calculate the radius profile, we divided the whole length of DNT into small segments of 1Å along the helical axis of DNT, next we find the center of each segment and calculate its distance from all the atoms in that segment. The average distance or radius of each segment from last 10 ns along with the standard deviation has been plotted against the segment position in figure 4b. We observe that the DNT maintain a constant radius ca. 2.3 nm in the central region embedded in the lipid bilayer membrane for all the salt concentrations. It is to be noted here that this radius value also includes the van der Waals radius of the DNA double helix (ca.1 nm) which has to be subtracted to get the pore radius. The systems with 0.5M salt solution, show small broadening of radius (ca. 0.5 nm) at both ends as compared to the central region whereas the system of 1M salt solution shows mostly a constant radius profile demonstrating extra stability at 1M salt concentration. In contrast, 0M system displays large widening in the radius profile of DNT on both ends. The radius profile and snapshots of the DNT lumen portray the gating like events similar to what was reported by Maingi *et al* in their recent study on DNA origami nanotubes in aqueous environment.(59) We find that the higher ionic concentration helps to keep the individual helical domains of DNT intact; hence they better maintain the tubular structure. At no salt (0M case), system does not have enough ions to screen the electrostatic repulsion between helical domains and as a result the individual helices connected via the Holliday junction, opens up and bend towards the lipid bilayer membrane. We see that the higher ions concentration also adds to the stability of DNT by providing more flexibility to the Holliday like crossover junctions in DNT.

**Ion Density Distribution and Dynamics**

It is known from the literature that depending on the size of cation, they possess different binding affinities and different binding sites to DNA.(60-62) The van der Waal radii of Na+, K+ and Cl- ions used in our simulation are 1.37 Å, 1.70 Å and 2.51 Å respectively. With a smaller radius, Na+ ions penetrate deeper into the grooves of DNA, condensate the structure more as compared to K+ ions. This process gives rise to the more stable DNT in lipid bilayer membrane in Na+. We observe that the equilibrium structure of DNA nanopores (DNPs) significantly varies for different salt concentrations. To explore the distribution of ions along the bilayer normal we have computed the average number density of cations and anions from the last 10 ns of 0.2 μs long equilibrium MD simulation as shown in figure S3a and S3b respectively of appendix S3 in SI. For all the simulated systems we see a dip in the ion density at the central region containing the impermeable membrane patch except the charge neutral case (0M system). For this system, we have fewer cations compared to other systems. Large fraction of these ions enters inside the DNT lumen to neutralize the charge of DNT backbone and give rise to a relatively flatter ion density profile. We have also calculated the diffusion coefficient of ions from the slope of mean square displacement (MSD) curves shown in figure S3b and S3c. Table 1 gives the values of the diffusion coefficient of various ions in different systems. The details of ion distribution and MSD calculations are given in appendix S3 of SI.

**Fluidity of Lipid Bilayer Membrane**

To quantify the changes in the structure of membrane due to the insertion of DNT, we have computed the time evolution of the area per head group (APH) available for the lipid molecules in each leaflet as shown in figure 4c. APH decreases initially (up to 50 ns) and then becomes flatter with small fluctuations around a constant value for rest of the simulation. Table 1 summarizes the values of APH averaged over last 10 ns of the simulation. It is observed that the system in NaCl solution shows significantly lesser APH as compared to KCl. We found that the NaCl salt solutions make the system more stable and compact. Due to the electrostatic interactions between zwitterionic lipid membrane and ions, the self-diffusion of the lipid molecules decreases with the ionic strength of the solution. Hence, 1M NaCl solution has lowest APH as compared to the all other simulated systems. Next, we examine the normalized electron density along the bilayer normal averaged over last 10 ns as shown in Figure 4d. This curve reflects various regions with different charge density in the system and shows the compact structure of the DNT embedded in the lipid membrane. The central region with two peaks and one cusp at the center corresponds to the bilayer thickness with DNT embedded. The peaks are slightly taller in 1M salt solution reflecting the compact bilayer. This region is followed by the sudden decrease in electron density which originates at the terminal of the lipid bilayer membrane. After this region, bare DNT expands in both directions. The thickness of the lipid membrane is grossly constant for the equilibrium structure in all salt concentration.

**Ionic Current and I-V Characteristics**

To calculate the transmembrane ionic current, we performed MD simulations at five different values of transmembrane voltage: 10 mV 20 mV, 50 mV, 100 mV and 200 mV. We took the equilibrated structure after 100 ns long MD simulation and apply the electric field in the direction of bilayer normal aligned along z axis. We performed the simulations with the electric field for 50 ns for all the cases (five different salt concentrations and each for five different values of transmembrane voltage). In order to maintain the structural integrity of the porin under the effect of electric field, the terminal atoms of DNT and heavy atoms of membrane were constrained to their equilibrium structure using a harmonic force constant of 0.1 kcal/mol.Å$^{-2}$. This restrain also prevents escaping of the DNT from the bilayer. The current density and hence the current of cylindrical nanopore can be estimated by equation 1 as follows

$$J = nqv_d = \frac{N}{AL_z}qv_d$$

$$\Rightarrow I(t) = JA = \frac{1}{\Delta t L_z}\sum_i^N q_i[z_i(t+\Delta t) - z_i(t)] \qquad (1)$$

Where N is total number of charge careers (ions) with drift velocity $v_d$, $z_i(t+\Delta t)$ and $z_i(t)$ are the z coordinate of the i$^{th}$ ion at the interval of current sampling frequency (2 ps), $L_z$ is the thickness of the membrane which is taken to be 5 nm and $q_i$ is the charge of the ion. We observed the steady flow of ions across the membrane which sets up a constant current for a given transmembrane potential. We calculated the ionic current for cations and anions separately and also calculate the total current. It is found that cations contribute to most of the total ionic current. Because of the electrostatic repulsion of the negatively charged DNT, Cl- ions do not prefer to enter inside the lumen giving rise to lesser contribution to current. We performed the running average of the total current values over 1 ns block. Figure S4a-e shows the ionic current as a function of the simulation time at five different molarities at various transmembrane voltage differences. Further, we averaged the ionic current values for last 10 ns of the simulation and plotted the average transmembrane current as a function of the voltage difference. Figure 5a shows the I-V characteristics of DNT lumen embedded in POPC lipid membrane at various salt concentrations. I-V curve shows that the channel is following the Ohm's law reasonably well. The solid lines show the linear fit for the I-V data points. The slope of the fit gives the Ohmic conductivity of the pore. The increasing molarity provides more charge careers, leading to the higher ionic conductance values. Table 1 summarizes the values of Ohmic conductance for the DNT lumen which vary from 4.31 nS to 20.64 nS depending on the salt concentration. Ionic conductance increases with increasing salt concentrations. The systems with KCl show larger conductance as compared to the system with NaCl due to the lower binding affinities K+ ions to DNA. This is also consistent with the ions diffusion analysis presented in previous section. It is important to note here that the ionic conductivities of the simulated systems are an order of magnitude higher compared to experimentally reported values from various groups as summarized in table 2. Conductance of 20.64 nS is by far the highest conductance reported for 6-helix DNPs in experiments. In fact, previously published simulation studies by Aksimentiev et al. also found higher ionic conductance for simulated DNPs as compared to the experimentally measured values.(14,26) The simulation results emphasizes that there is a scope of improvement in the experimental understanding of the ionic conductance through membrane protruding DNPs. We believe that the equilibrium orientation of the DNPs achieved through their self-assembly in experimental synthesis could be different than what is in general presumed and implemented in simulation studies. We speculate that the equilibrium orientation of the DNPs achieved through their self-assembly in experimental synthesis may be different than what is in general implemented in simulation studies. In simulation, we start with an ideal configuration of DNT embedded inside the hydrophobic core of the bilayer and perfectly aligned to bilayer normal. This could be different than what is synthesized by the self-assembly of nucleotides in the experiments. Further, a detail and careful study would be required to assess the fidelity of DNPs in lipid bilayer membranes and address the anomaly of lower ionic conductance measured in experiments. To get a theoretical estimate of this conductance we treat the DNT as a nanopore for a constant electrolyte transport. The conductance of such a cylindrical nanopore with channel length $l$, pore diameter $d$ can be modeled using the following equation.

$$G = \kappa \left\{ \frac{4l}{\pi d^2} + \frac{1}{d} \right\}^{-1} \quad (2)$$

where $\kappa$ is the bulk conductance of electrolyte solution Here the second term in the denominator is the correction due to the hemispherical geometry of the nanopore at both the ends, which is also referred as the access resistance. (63) This simple model is known to correctly fit a wide range of experimental data from synthetic membrane nanopores.(64) Using the typical values of conductance for the bulk solution of 1M NaCl (8.5 S/m) and 1M KCl (10.86 S/m) at normal temperature (65) in equation 2, we obtain a theoretical conductance of 1.27 nS and 1.62 nS respectively. In this theoretical framework, we approximated the DNTs with fix radius of 2 nm. But from the radius profile as shown in figure 4b, we know that the DNTs swell during the course of simulation to attain the equilibrium structure. As the porous nature of DNT has been ignored in the theoretical estimation, this could be also a reason behind the lower conductance vales.

**Ligand Gated DNPs**

Controlled permeation of molecular cargo across the biological membrane is extremely relevant in biotechnology. Recently, Burns et al. synthesized DNPs which can be open and closed by the hybridization of the 'lock' and 'key' strands at mouth of the channel.(16). It is interesting to see if such gating mechanism can be probed in simulation also. The kinetics of hybridization is a slow process and to the best of our knowledge has not been demonstrated in all atom MD simulations. To model the gating behavior of DNPs, we design the all-atom model of nanopores conjugated with two loops of single stranded DNA (ssDNA) overhangs at the mouth DNTs which we term as closed state (NPC) and with hybridized ssDNA to its complimentary strand which we term as open (NPO). These gated nanopores (NPC and NPO) are derived by adding the docked oligonucleotides to template 6-helix DNP (NP) in a similar manner as shown by Burns et al. in their supplementary figure 1.(16) Next, we perform extensive MD simulations of these closed and open state nanopores embedded in POPC lipid bilayer membrane in 1M electrolyte solution of KCl using the same simulation methodology discussed in the method section. The details of the model building and MD simulation are discussed in table S1 and appendix S5 of SI. Figure S5a and S5b show the snapshots of initial built and equilibrated structures of NPC and NPO after 100 ns MD simulation respectively. The analysis of the MD simulation trajectories shows that these transmembrane nanopores are stable in ambient conditions. In order to see the gating behavior due to the molecular design of NPC and NPO, we have computed the transmembrane ionic current at various transmembrane voltage differences as shown in figure S5a and S5b. From the figure we find that, for a given value of the transmembrane voltage difference, NPO always has higher ionic current as compared to NPC. From the average ionic current values shown in figure S5a-b, we have plotted the I-V characteristics curve for NPO, NPC and NP in figure 5b which shows Ohmic behavior of the simulated ionic currents. From the linear fitting of the I-V curve, we extract the ionic conductivity of these porous channels which comes out to be 27.53 nS, 19.27 nS and 20.64 nS for NPO, NPC and NP respectively. In case of the NPC, we have single stranded loops over the native 6-helix DNT. These flexible single strands (each having a length of 14 bp) come close to each other during the MD equilibration to adopt a stable in-solution structure due to solvent effects. In the process, they block the passage of ions and act as a lock to central ionic channel. On the other hand, in case of NPO, by the virtue of its design, these single stranded oligomers are hybridized with complimentary strands. Hence the oligomers stay in their helical domain and allow the uninterrupted passage of electrolyte solution in transmembrane channel resulting higher current. The top view of equilibrium snapshots of the simulated structures shows that the central channel is partially blocked in NPC due to the conformational fluctuation of the overhangs.

**CONCLUSIONS**

In this study, we demonstrate the stability and dynamics of a tile-based 6-helix DNT embedded in POPC lipid bilayer using the analysis of 0.2 µs long equilibrium MD simulation trajectories. The lipid molecules in vicinity of the DNT reorient themselves to form a toroidal structure around it. We propose that this mechanism reduces the energy barrier to self-assemble a porous lumen of DNT in lipid bilayer membrane. The spontaneous formation of a toroidal pore cost energy penalty to the system. In our previous work we have shown that the activation barrier for a toroidal pore formation in POPC bilayer is ca. 25-27 kJ/mol under an electric field strength of

0.3-0.8 V/nm.(56) At zero biasing potential, this could be substantially larger. As reported by Gopfrich et. al., a toroidal pore of 2nm radius costs an energy penalty of ca. 100-200 kJ/mol without any cholesterol anchor which reduces to -1000 kJ/mol for a pore of 6nm radius with 20 cholesterol anchors.(14) Hence, to embed DNT porin in the bilayer, the energy barrier can be reduced by the addition of cholesterol anchors to DNA. Currently, there is no study on the microscopic understanding of the mechanism of DNT insertion in the bilayer. All the existing simulation studies report structure and conductivities of DNA porin which is pre-inserted in the bilayer. A thorough study is required to understand the mechanism of insertion and the self-assembly process of DNT in bilayer. We also show that higher salt concentration makes the DNT more compact and helps to stabilize them better in the lipid environment. Further, we observe the higher stability of the DNT in presence of $Na^+$ as compared to $K^+$ counterions due to their better condensation. The ionic current measurement under constant electric field allows us to calculate the I-V characteristics of these pores. The system displays nearly Ohmic behavior over a wide range of transmembrane voltage and ionic concentration. The conductance of the DNT varies from ca. 4nS to ca. 21 nS depending on the ionic concentration. A conductance of ca. 21 nS is by far the highest conductance reported in literature for a DNP with radius of 2 nm. The MD simulations with open and close pore models of DNT capture the molecular gating of these ion channels. The results from this all atom MD simulation study will help to understand the phenomenology of DNPs in lipid bilayer membranes which will lead to their rational design for evolutionary biophysical applications.

**Acknowledgement**

We thank DAE, India for financial support. We acknowledge Supercomputer Education and Research Center, (SERC) IISc Bangalore for providing access to the high performance supercomputer SahasratT. H.J. thanks to CSIR, GoI for research fellowship. We acknowledge Prof. Ned Seeman for the extremely helpful inputs about the design of DNTs. We thank Prof. Arnab Rai Choudhuri for useful discussions.

**Supplementary Data**

Supplementary Data are available at NAR online.

**Competing financial interests:**

The authors do not declare any competing financial interests.

**Figures and Table**

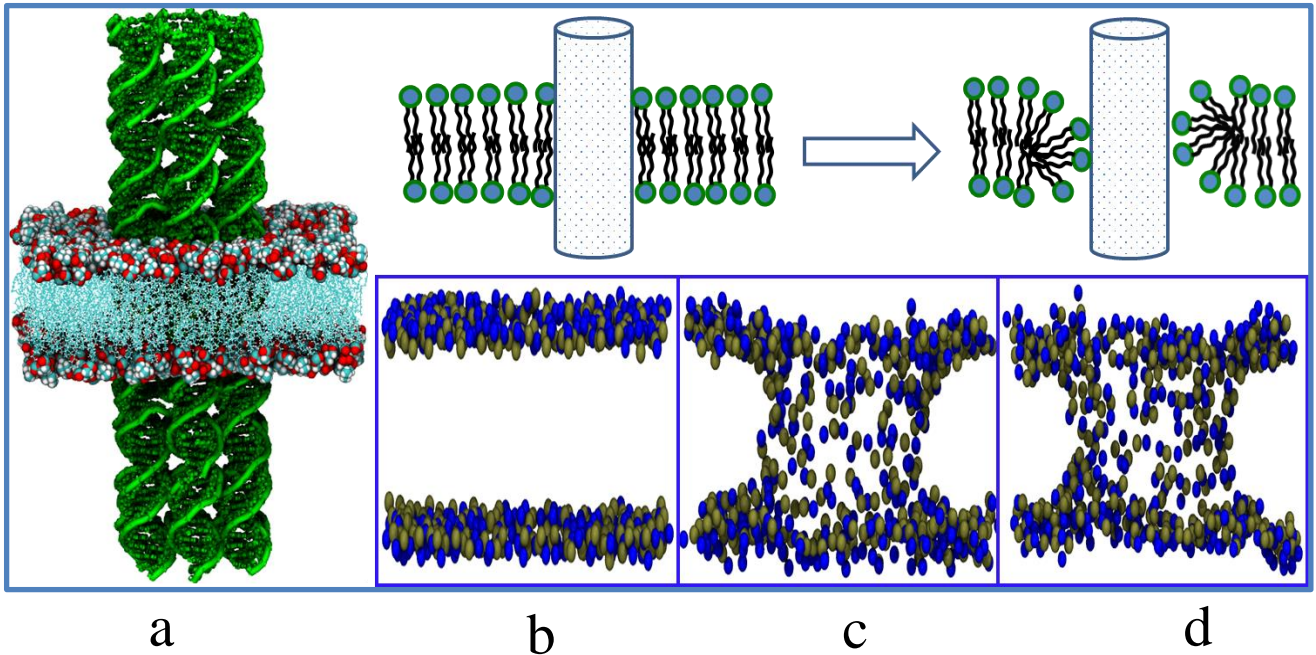

a   b   c   d

**Figure 1:**

(a) Initial built structure of DNT embedded in lipid bilayer membrane. Snapshots of Nitrogen (blue) and Phosphorous (yellow) atoms of lipid head groups; (b) at the beginning of the simulation (0 ns), (c) after 15 ns and (d) after 205 ns of equilibrium MD simulations of lipid bilayer membranes. The snapshots show that the lipid head group atoms tilt towards the DNT and form a toroidal structure around it within few nanoseconds, Further, they maintain this arrangement throughout the entire course of MD. Top panel above the figures shows the schematic representation of the lipid head group reorientation.

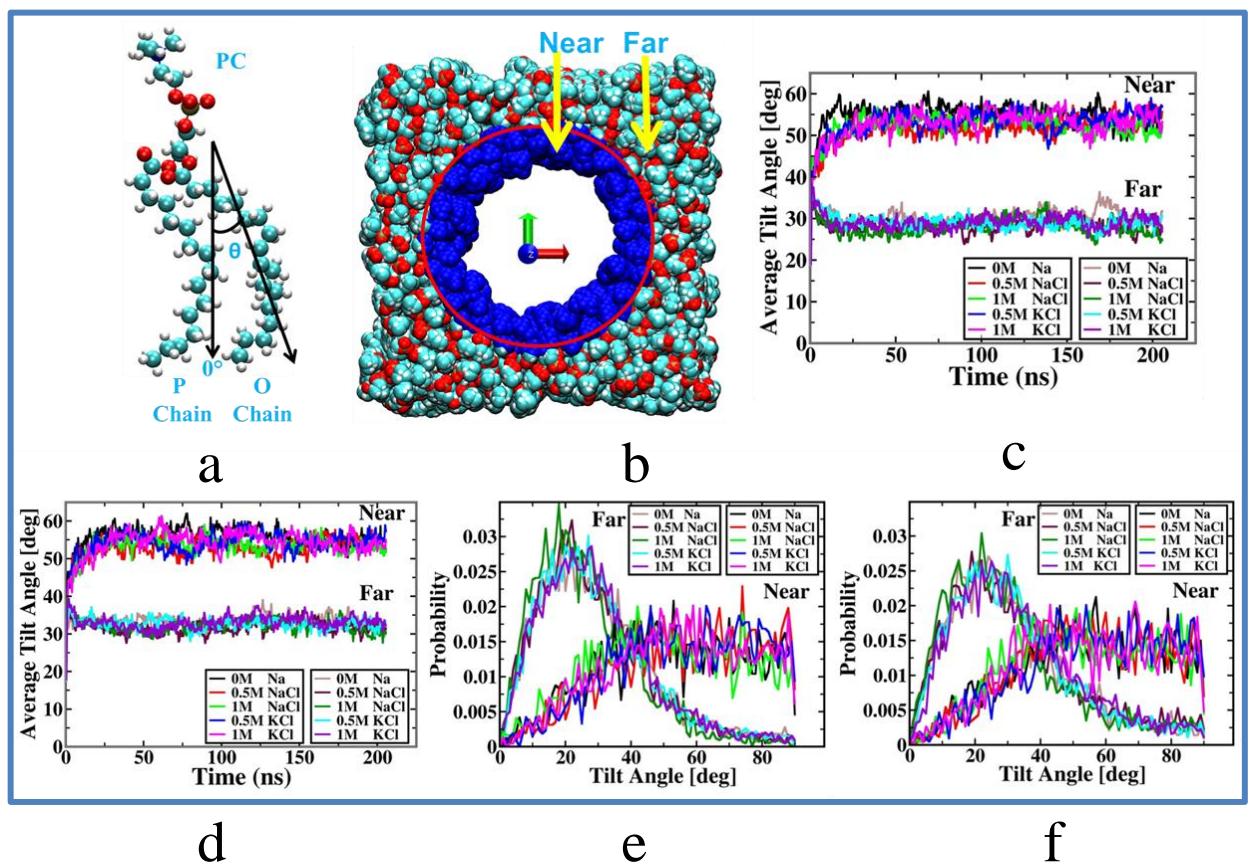

**Figure 2:**

(a) The definition used to calculate the lipid tilt angle, (b) convention of far and near lipids. Evolution of tilt angle for far and near lipid chains as a function of the simulation time (c) for Palmitoyl (P) and (d) for Oleoyl (O) chains respectively. Probability distribution of tilt angles averaged over the last 10 ns of 205 ns long simulations (e) for P and (f) O chains respectively. The analysis shows that tail lipid chains tilt away from the DNT channel and form a toroidal structure around it.

a b c d e

**Figure 3:**

The instantaneous snapshots of the DNT embedded in lipid bilayer membrane (excluding water) after 205 ns equilibrium MD at various salt concentrations (a) 0M, (b) 0.5M NaCl (c) 1M NaCl, (d) 0.5M KCl and (e) 1M KCl. The top panel shows the view of the system along the bilayer normal and bottom panel shows the side view. For the sake of clarity, water molecules are not shown.

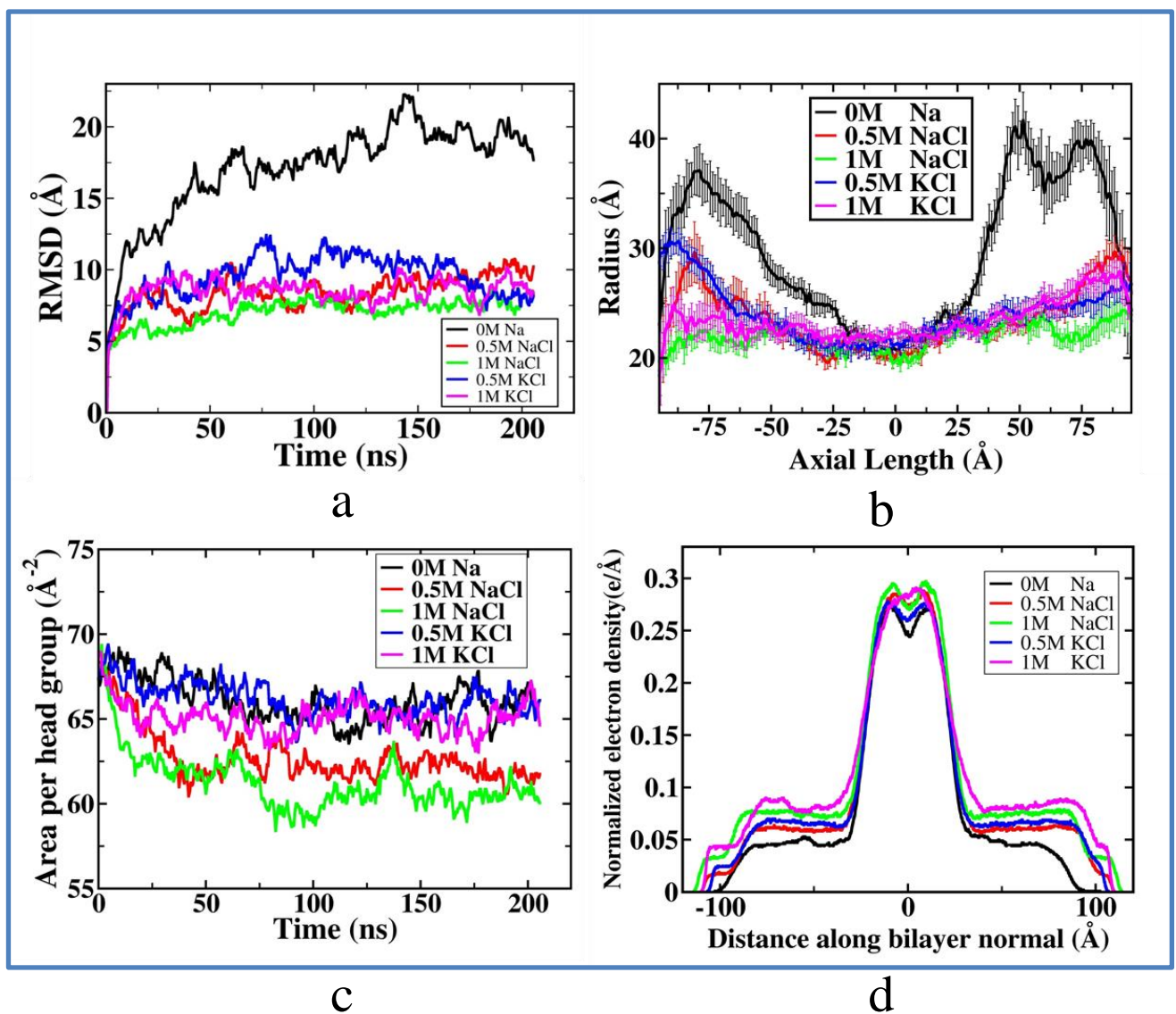

**Figure 4:**

(a) Time evolution of RMSD of the DNT with respect to the energy minimized structure. Higher salt concentration keeps the DNT structure better stabilize and more compact in lipid environment and RMSD decreases with increase in salt concentration. (b) The radius profile of the DNT along its helical axis. The radius is averaged over the snapshots from last 10ns of the MD simulation. DNT with 0M salt concentration is not maintaining its tubular structure except the central region extending in bilayer, where it has a uniform radius of 2.2 nm. System with higher salt concentration has a uniform radius profile throughout the nanotube length. (c) Time evolution of the area per head group of the lipid bilayer membrane. (d) The electron density profile of the system along the bilayer normal averaged over the snapshots of last 10 ns MD simulations. The center region of this curve extending from -20 to 20 Å reflects the width of lipid bilayer with a cusp showing the middle of the membrane and the peaks corresponds to the polar head groups followed by the dip representing DNT with the counterions.

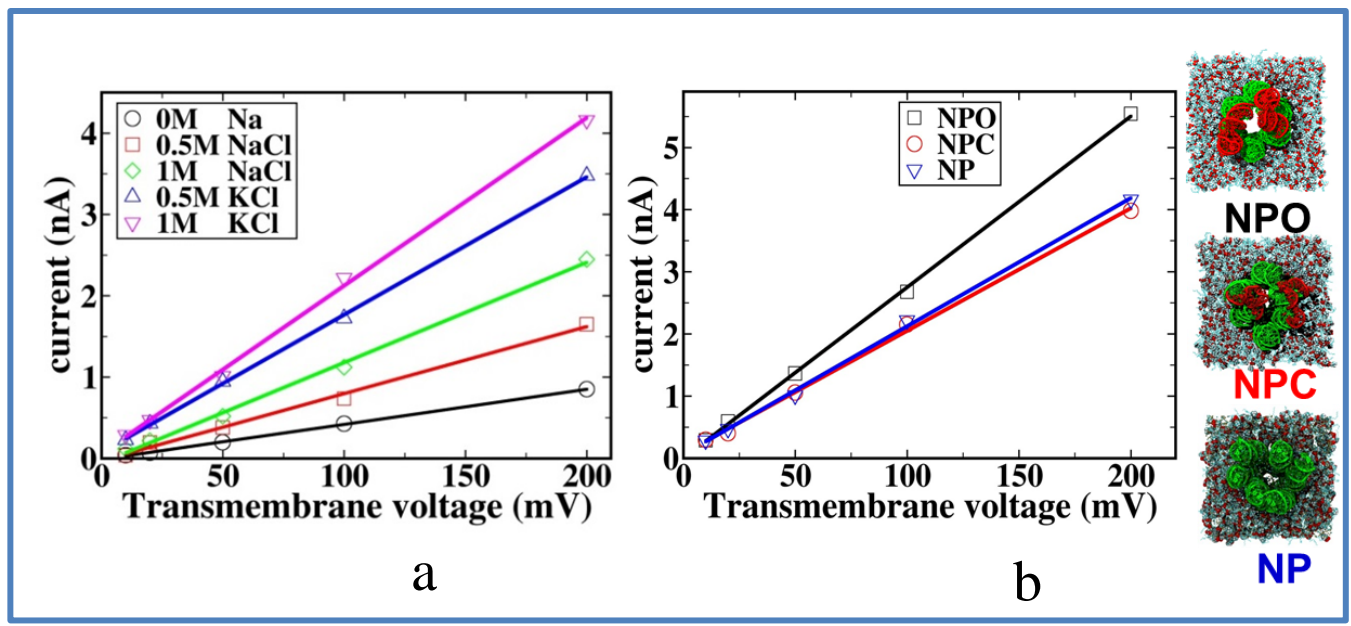

**Figure 5:**

(a) Simulated transmembrane ionic currents through the DNT as a function of different transmembrane voltages difference for different salt concentration. The solid line is the linear fir to I-V data and gives the Ohmic conductivity of the system. (b) Similar, I-V characteristic curve for NPO, NPC and NP, the top view of respective structure after 100 ns equilibrium MD simulation is shown on the right. NPO shows significantly high ionic conductivity as compared to NPC and NP which have almost similar conductance values.

# Tables

**Table 1: Summary of the Analysis.**

Area per head-group, average tilt angle for lipid chains and diffusion coefficient of the ions during the equilibrium MD simulations for the system at various salt concentration. The APH and tilt angle has been averaged for last 10 ns of MD simulation. The last column provides the Ohmic conductivity of the channel derived from I-V characteristics curves.

|  | 0M Na | 0.5M NaCl | 1M NaCl | 0.5M KCl | 1M KCl |
|---|---|---|---|---|---|
| **Area per head group ($Å^2$)** | 66.3 ± (0.5) | 61.6 ± (0.3) | 60.6 ± (0.3) | 65.6 ± (0.4) | 66.1 ± (.64) |
| **Average Tilt Angle (°) (P chain) Near/Far** | 52.7/ 29.6 ± (1.3/1.2) | 56.0/28.5 ± (0.9/1.4) | 50.6/26.1 ±(1.2/1.0) | 55.7/29.2 ±(1.6/1.1) | 54.1/29.3 ±(1.7/1.2) |
| **Average Tilt Angle(°) (O chain) Near/Far** | 54.2/31.3 ±(1.7/1.1) | 54.3/32/2 ±(1.5/1.1) | 52.9/30.0 ±(1.4/1.0) | 55.8/31.8 ±(1.3/1.1) | 53.9/33.2 ±(1.1/1.1) |
| **Diffusion coefficients $Na^+/K^+$ ions ($10^{-9} m^2 s^{-1}$)** | 0.53 | 0.85 | 0.80 | 1.44 | 1.38 |
| **Diffusion coefficients $Cl^-$ ions ($10^{-9} m^2 s^{-1}$)** | - | 1.62 | 1.33 | 1.77 | 1.56 |
| **Conductivity (nS)** | 4.31 | 8.24 | 12.32 | 16.95 | 20.64 |

**Table 2: Summary of the conductance values obtained from various studies.**

| Reference Article | Conductance (nS) | Theoretical Radius (nm) | Molarity and Salt type | Nature of DNP |
|---|---|---|---|---|
| Burns et. al. Nano Letter 2013[10] | 0.395 | 2 | 1 M KCl | 6HB Origami |
| Langecker et. al. Science 2012[11] | 1.0 | 2 | 1 M KCl | 6HB Origami |
| Göpfrich et.al. Nano Letter 2016[13] | ~0.5 - 1.5 | 1.6 | 1 M KCl | 4HB DNA-Tile |
| Gopfrich et.al. ACS Nano 2016[14] | 30 (Exp) 46 (MD) | 7.5 (Square) | 1 M KCl | Funnel-Shaped Square Origami |
| Krishnan et.al. Nat. Comm. 2016[15] | 3 | 4.5 (Square) | 1 M KCl | Square Origami |
| Gopfrich et.al. Nano Letter 2016[25] | 0.1 (Exp) 0.095(MD) | - | 1 M KCl | Duplex DNA |
| Yoo et.al. JPCL 2015[26] | 6.9 | 2 | 1.4 M KCl | 6HB Atomistic Model |
| Current Study | 20.64 | 2 | 1.45 M KCl* | 6HB Atomistic DNA-Tiles |

* The molarity of the system increases from 1M to 1.45 M during the course of initial 5 ns equilibrium NPT simulation as the volume of the simulation box shrinks and attains the normal density.

# Supporting Information

**1. Simulated System Details:**

Figure S1 shows the building the structure of DNT embedded in lipid bilayer membrane at various stages. We first create a 3nm circular pore in POPC lipid bilayer and then align the DNT to the center of the nanopore as shown in Figure S1a. Next we solvate the structure in water box and add ions as per the respective salt concentration.

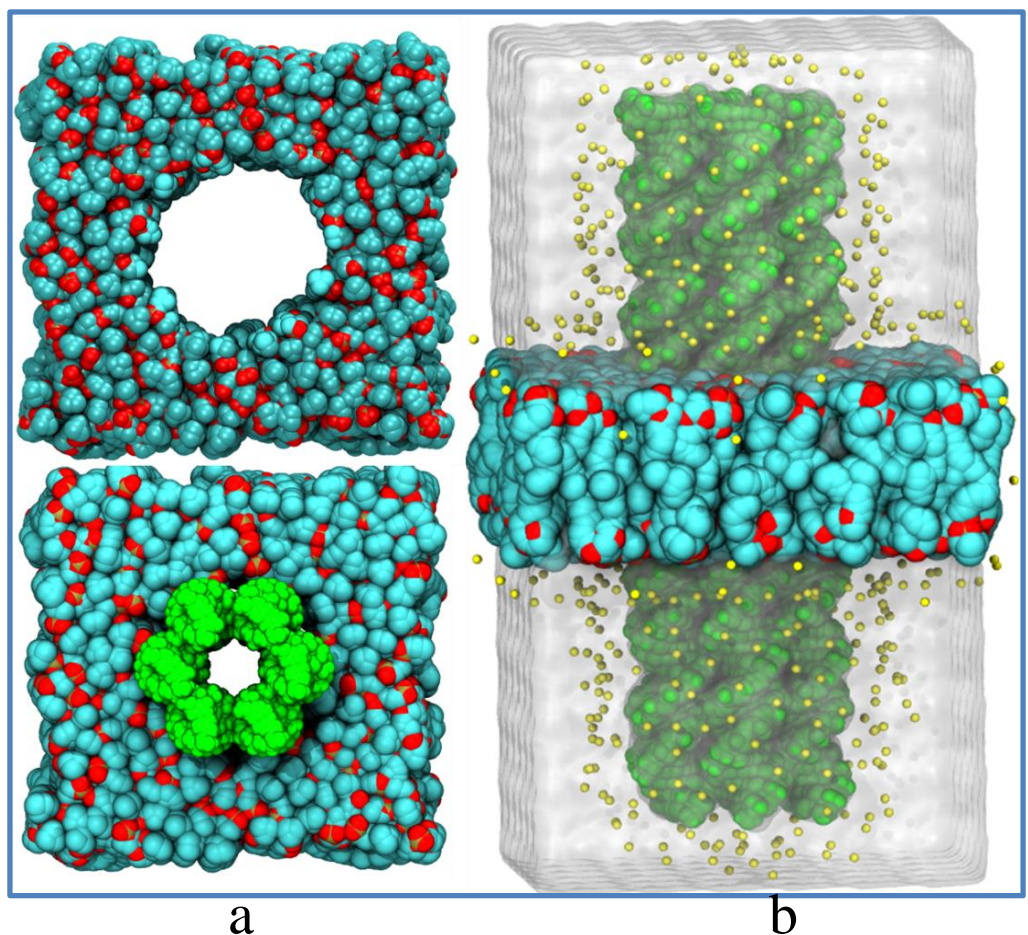

Figure S1: (a) (Top panel) The atomistic model of POPC lipid bilayer membrane nanopore with a diameter of ~5 nm. (Bottom panel) Top view of the initial built structure of the DNT embedded in lipid bilayer membrane. (b) The side view of the system after solvating into a TIP3P water box and charge neutralized with counterions. DNT, lipid and ions are shown in vdW representation with different colors whereas the water molecules are represented as a continuous white background. The system comprises ~0.3 million atoms.

Each system contains a DNT (21691 atoms,) embedded in a POPC lipid bilayer membrane (45024 atoms) along with water molecules and the monovalent ions. The dimension of the periodic box are [138 137 232] Å for every system at the beginning of the simulation. Based on the volume of the simulation box, we compute the number of ions to be added to achieve the desired molarity of the solution. As the simulation proceeds the box size reduces which leads to higher molarity of the system.

Table S1: Atomistic details of the Simulated Systems

| System | Total number of Atoms (DNT+POPC+Ions+WAT) | Na+/K+ Ions | Cl- Ions |
|---|---|---|---|
| 0M Na | 297875 | 664 | 0 |
| 0.5M NaCl | 298930 | 1985 | 1321 |
| 1M NaCl | 300428 | 3307 | 2643 |
| 0.5M KCl | 299392 | 1985 | 1321 |
| 1M KCl | 301082 | 3307 | 2643 |
| NPC (1M KCl) | 374071 | 3993 | 3257 |
| NPO (1M KCl) | 397953 | 4079 | 3257 |

* NPC and NPO are the atomistic models of open and closed state nanopores which contain 24430 and 26924 atoms of DNA respectively.

## 2. Evolution of the 0M system during the 0.5 μs MD simulation.

In order to see the temporal stability of DNA nanotube (DNT) embedded lipid bilayer membrane system over longer time scales, we have performed 0.5 μs long equilibrium MD simulations of 0M system using the similar simulation methodology discussed in this paper. The analysis of the simulation trajectory shows that the DNP is intact and stable over this microsecond long equilibrium MD simulation. Figure S2a and S2c show the snapshot of the system at the beginning and after 0.5 μs equilibrium MD simulation respectively. To see the diffusion of DNT across the membrane we have plotted the time evolution of the lateral distance between center of mass of DNA nanotube and lipid bilayer along the bilayer normal (z direction) in figure S2b. This plot conclude that the DNP is stable in membrane without any hydrophobic modification due to the lipid reorientation which shields the hydrophobic tails of lipid from the water filled lumen of DNA nanotube. The simulation snapshots also show the considerable bending of DNT outside the membrane patch towards the lipid bilayer membrane in 0M system. The deformation is also reflected in the higher RMSD values as shown in figure S2d. This is due to the electrostatic interaction of lipid head groups with the charged DNT. However, as we increase the salt concentration the tubular structure is well maintained as this electrostatic interaction is screened. As the simulation evolves, the lipid rearrangement compensate for the energy cost of inserting the highly charges DNT in lipid membrane giving rise to a toroidal structure around the water filled lumen. Although we start with an energetically unfavorable conformation but once the toroidal pore is formed it can host the membrane spanning DNA nanopores as confirmed by various structural analyses performed in our study. Previously, Khalid *et al.* (57) have shown that there exists a large free energy barrier (~ 40 kJ/mol) to insert the DNA inside the lipid bilayer membrane. This barrier can be suppressed by hydrophobic modification(27) but as Khalid *et al.* concluded that the lipid rearrangement can also potentially suppress this barrier which is exactly what we observe in this study.

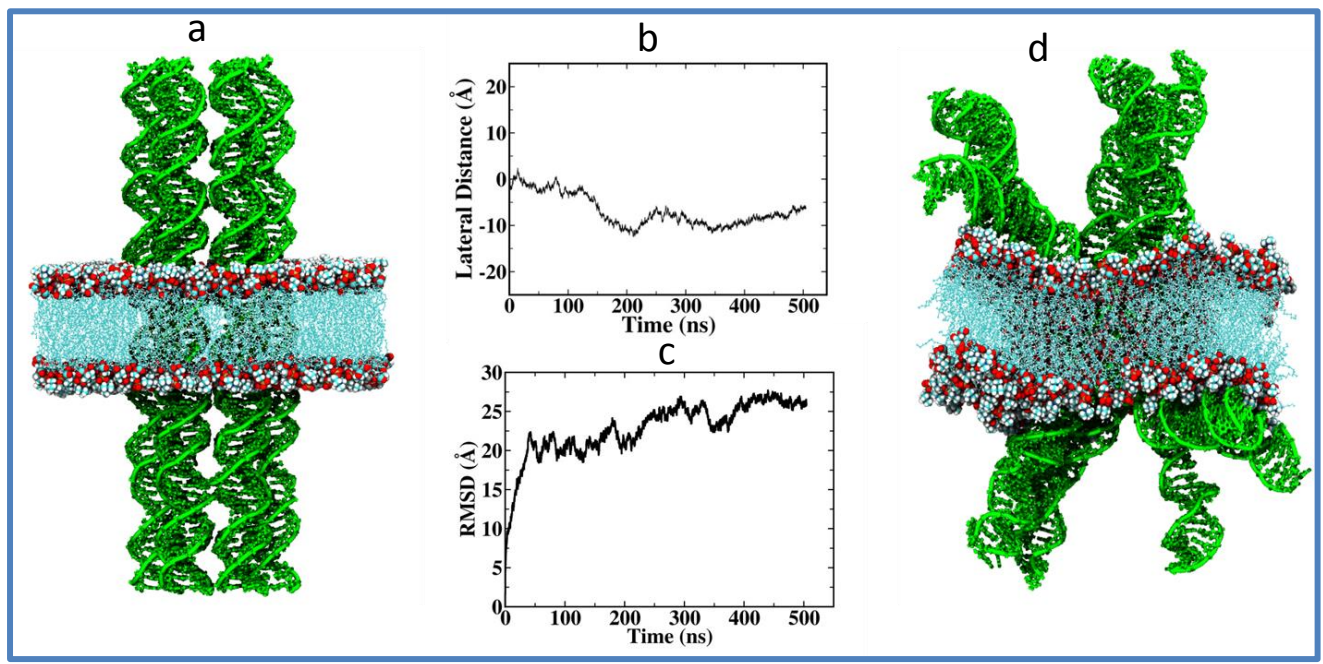

*Figure S2: (a) Initial structure of DNT embedded in lipid bilayer membrane. (b) Time evolution of the lateral distance between the center of mass of DNT and lipid bilayer membrane along the bilayer normal (z direction) during the course of 0.5 µS long equilibrium MD simulation. (c)Evolution of RMSD of DNT with respect of the initial energy minimized structure (d) The snapshots of the structure after 0.5 µS simulation.*

### 3. Ionic Density along Bilayer normal and Mean Square Displacements.

As we have seen striking molarity dependence on the structure and stability of DNT embedded in lipid bilayer membrane, we further analyze the behavior of ions during the simulation. Figure S3a and S3b show the number density profile of the cations (Na+/K+) and anion (Cl) along bilayer normal. The density has been averaged over last 10 ns of the MD simulation. For all the simulated systems we see a dip in the ions density at the central region containing the impermeable membrane patch except the charge neutral case (0M system). For this system, we have least cations compared to as other systems, here large fraction of these ions enters inside the DNT lumen to neutralize the charge of DNT backbone and give rise to a relatively flatter ion density profile (Figure S3a). To quantify the movements of ions in the system, we have also calculated the Mean-Square-Displacements (MSD) for ions using the following formula,

$$(\Delta r^2(t)) = \frac{1}{N} \sum_{i=1}^{N} <r_i(t+t') - r_i(t)]^2>_{t'} \quad (S1)$$

where t' is the time origin and N is total number of ions.

To get better statistics for diffusion, we extracted 2 ns MSD data from 205 ns MD simulation trajectories using time origin averaging method. Previously, we have used similar protocol to calculate the diffusion coefficients of confined water.(66) Figure S3c and S3d shows time origin averaged MSD of cations (Na+/K+) and anions (Cl-). From the linear fit of the MSD plot, we extracted the diffusion coefficients of ions using the Einstein relation. The diffusion coefficient of Na+ ions turns out to be 0.53 x$10^{-9}$ $m^2s^{-1}$, 0.85 x$10^{-9}$ $m^2s^{-1}$ and 0.80 x$10^{-9}$ $m^2s^{-1}$ at 0M, 0.5M NaCl and 1M NaCl salt concentration respectively. Similarly, for K+ ions the diffusion coefficient is 1.44 x$10^{-9}$ $m^2s^{-1}$ and 1.38 x $10^{-9}$ $m^2s^{-1}$ at 0.5M KCl and 1M KCl salt concentration respectively. Due to the binding of counterions to negatively charged DNT backbone and impermeable lipid membrane patch, the diffusion coefficients of the counterions are understandably lower compared to the their bulk diffusion coefficients at as reported by Varnai *et al.* 1.7 x $10^{-9}$ $m^2s^{-1}$ for Na+ ions and 2.8 x $10^{-9}$ $m^2s^{-1}$ for K+ ions.(60) For Cl- ions, the diffusion coefficient is 1.62 x$10^{-9}$ $m^2s^{-1}$, 1.33 x$10^{-9}$ $m^2s^{-1}$, 1.77 x $10^{-9}$ $m^2s^{-1}$ and 1.56 x$10^{-9}$ $m^2s^{-1}$ for 0.5M NaCl, 1M NaCl, 0.5M KCl and 1M KCl respectively. The diffusion coefficients for the Cl- ions are higher as compared to the cations. This is expected due to the electrostatic repulsion of negatively charged DNT backbone to Cl- ions. It is important to note that K+ ions diffuse faster through the DNT as compared to the Na+ ions. The slower diffusion Na+ ions is due to the tight binding of Na+ ions to DNT backbone as compared to K+ ions as discussed in the previous section. Hence K+ ions have higher mobility and greater diffusion coefficient as compared to Na+ ions through DNT. The values of diffusion coefficient for the ions in 1M solution systems are lesser than the corresponding values in 0.5M. This is because of the steric hindrance in 1M systems with excessive cations and anions.

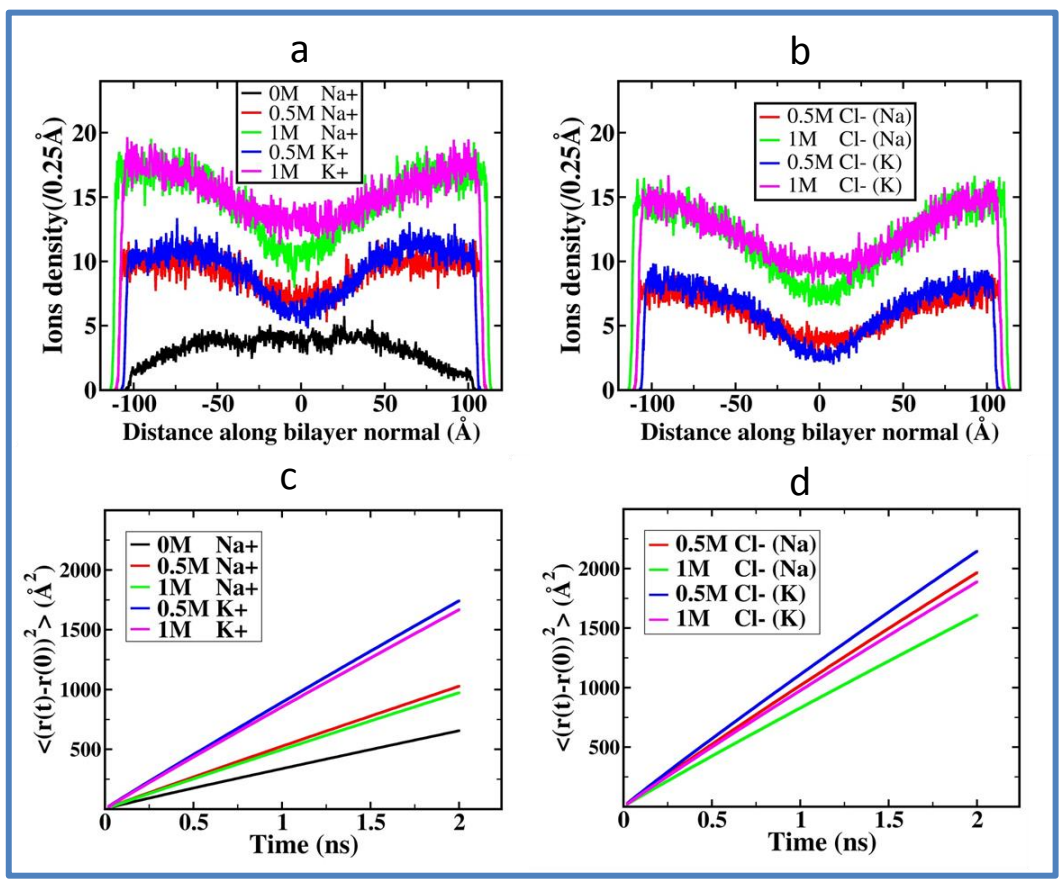

Figure S3: Ion distribution map averaged over last 10 ns of the MD trajectories along the axis of the DNT (a) for cations (Na+/ K+) and (b) for anions (Cl-). The time origin averaged MSD of (c) positively charged ions (Na+/K+) and (d) negatively charged Cl- ions. The diffusion of negative ions is higher due to negatively charged DNT. The respective values of diffusion coefficients have been shown in table 1.

## 4. Transmembrane Ionic Current.

To trace the I-V characteristics of DNPs in MD simulations, we have performed 5 simulations at different transmembrane voltage difference for each structure. We run these constant electric field simulations for 50 ns. Using the formula given in equation 1 of the main manuscript, we computed the transmembrane ionic current for each simulation. Figure S4 a-e shows the ionic current as a function of simulation time for 0M, 0.5 M NaCl, 1M NaCl, 0.5 M KCl and 1 M NaCl respectively.

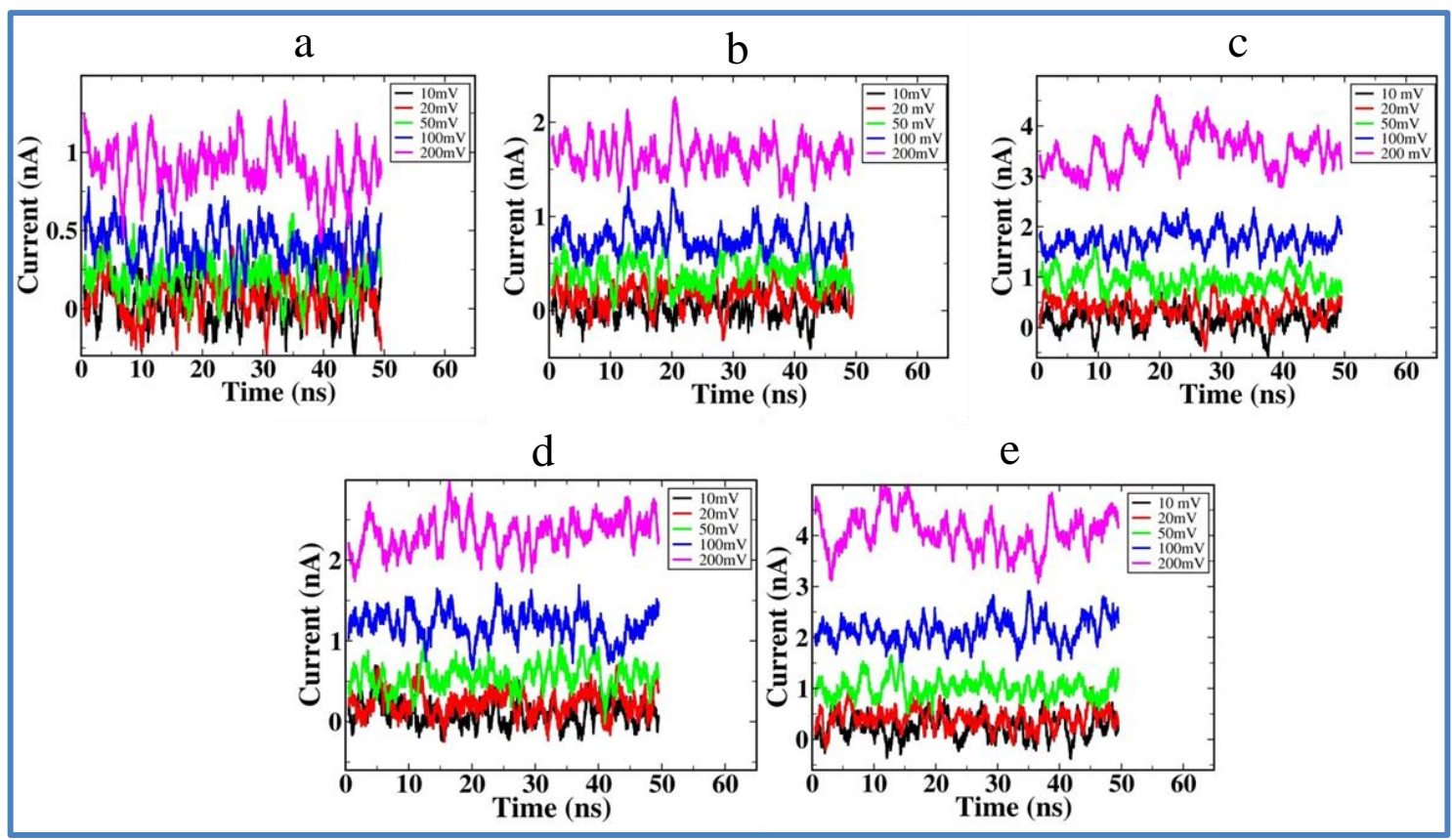

Figure S4: Ionic currents through the DNT lumen in lipid membrane during the simulation at different transmembrane voltage and varying salt concentration: (a) 0M (b) 0.5M NaCl, (c) 1M NaCl (d) 0.5M KCl and (e) 1M KCl.

## 5. Ligand Gated DNPs

To analyze the gating events in membrane spanning DNPs by conjugating the docked oligonucleotide to the template 6-helix DNT as suggested by Burns et al,(16) we have built and simulated the atomistic models of closed and open state DNPs. Figure S5a and S5b shows the atomistic models of NPC and NPO respectively. The template 6-helix DNT is shown in green color and the conjugated docked oligomers attached for the purpose of controlled gating are shown in red color. To generated NPC, we added two loops of ssDNA at the mouth original 6-helix DNT. As the structure evolves in simulation, these flexible ssDNA overhangs acts as molecular lock by coming closer and partially blocking the ions flow into the channel.

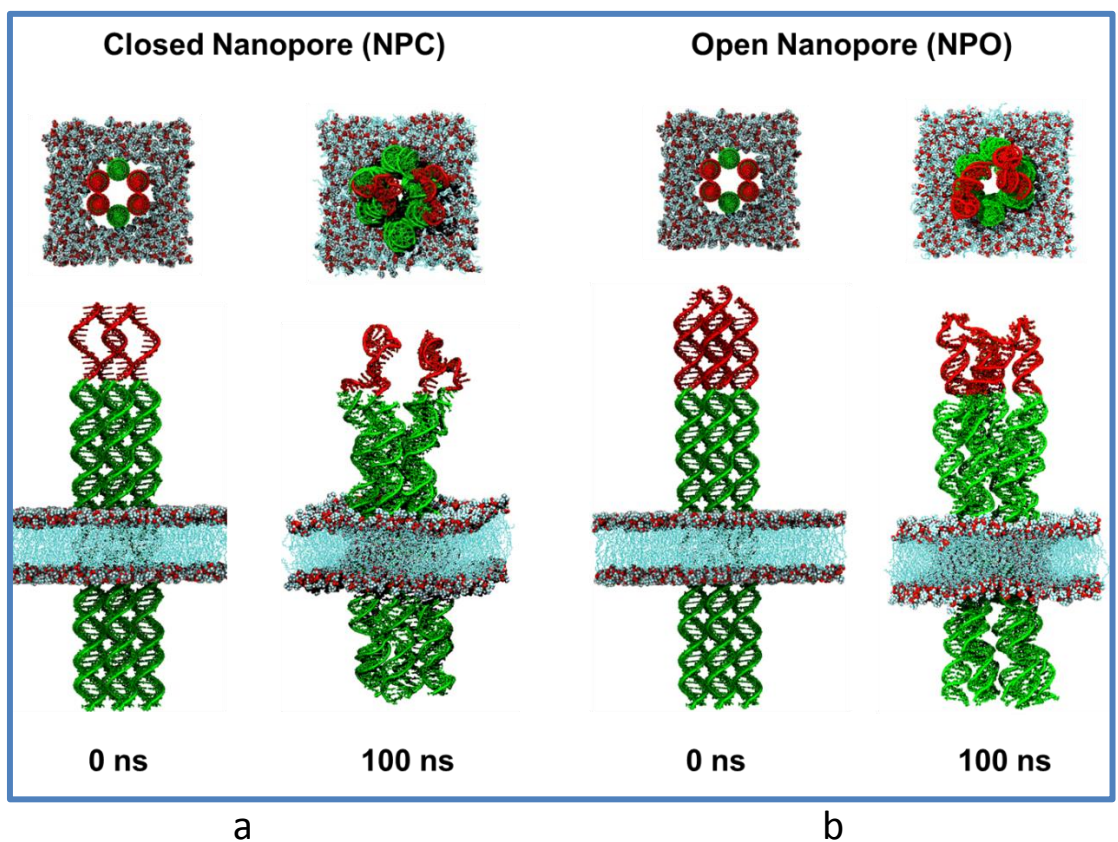

Figure S5: *Snapshots of the atomistic simulations of ligand controlled DNPs,(a) NPC and (b) NPO, proposed by Howorka et al. for the controlled transport of cargo across the membrane. The template nanopore is shown in green color whereas the red color strands shows the docked oligonucleotides attached to achieve the desired molecular gating. Top panel shows the image of the respective structure along the bilayer normal. Water and counter-ions are not shown for the sake of clarity.*

To design NPO, we modify the ssDNA of NPC by hybridizing them with their complimentary sequence and making the double helical DNA as shown in red color in figure S5b. These double helical domains in NPO are stiff as compared to the ssDNA domain of NPC and allow the free passage of ions in the channel. Table S1 summarizes the atomistic details of various constituents of NPC and NPO. To compare the ionic permeability of NPC and NPO, we perform MD simulation at constant transmembrane voltage differences. In figure S6a and S6b, we plot the transmembrane ionic current recoding of NPC and NPO at different values of transmembrane voltage differences along with the simulation time. The plot confirms that, for a given value of transmembrane voltage difference, NPO always has higher ionic current as compared to NPC. The ionic current shown in the figure is averaged over a block of 1 ns which is fairly constant. Last 10 ns of data is averaged to get a current corresponding to a particular voltage difference and plotted in figure 5b of the main manuscript. We observe Ohmic behavior of transmembrane ionic current which allow us to extract the ionic conductivity of the nanopre from the slop of I-V characteristic curve. The larger value of ionic current for NPO with respect to NP is due to the more negative charge density of 14 bp long loops of dSDNA at the mouth of nanopore. Therefor NPC attract more K+ ions to the channel as compared to NP which leads to higher ionic current.

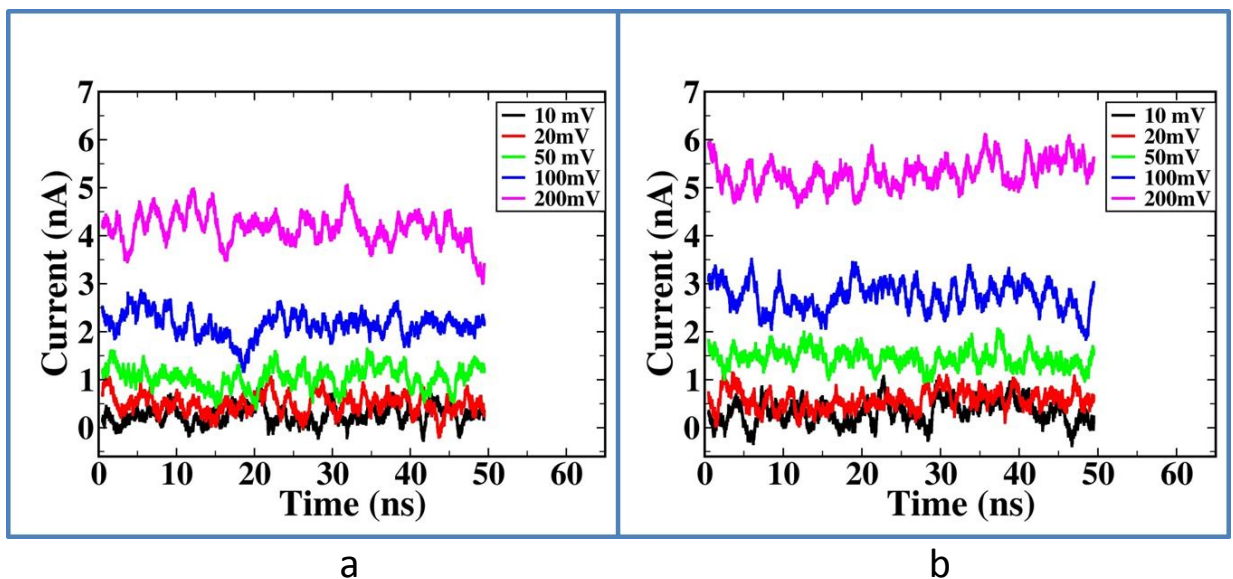

*Figure S6: Traces of transmembrane ionic current for gated DNPs in 50 ns MD simulation; (a) NPC, (b) NPO at various values of constant transmembrane voltage differences.*